\documentstyle[aps,preprint,multicol,epsf]{revtex}
\begin{document}

\title{Projective Measurement Scheme for Solid-State Qubits}
\author{Lin Tian$^{\mbox{\scriptsize a}}$, Seth Lloyd$^{\mbox{\scriptsize b}}$ and T.P. Orlando$^{\mbox{\scriptsize c}}$}
\address{$^{\mbox{\scriptsize a}}$Department of Physics, $^{\mbox{\scriptsize b}}$Mechanical Engineering, $^{\mbox{\scriptsize c}}$Electrical Engineering and Computer Science,
Massachusetts Institute of Technology}
\date{\today}
\maketitle

\begin{abstract}
We present an effective measurement scheme for the solid-state
qubits that does {\bf not} introduce extra decoherence to the
qubits until the measurement is switched on by a resonant pulse.
The resonant pulse then maximally entangles the qubit with the
detector. The scheme has the feature of being projective,
noiseless, and switchable. This method is illustrated on the
superconducting persistent-current qubit, but can be applied to
the measurement of a wide variety of solid-state qubits, the {\bf
direct} detection of the electromagnetic signals of which gives
poor resolution of the qubit states.
\end{abstract}


Quantum computation in solid-state systems is a growing field.
\cite{rochester_qubit,divincenzo_qubit,pc_qubit_s,pc_qubit_l,charge_qubit_1,charge_qubit_2,dwave_qubit,kane_qubit}
Among various physical realizations, solid-state qubits have the
advantage of being scalable to large number of qubits and that the
quantum states can be engineered by various techniques. Successful
implementations of qubits have been achieved in several mesoscopic
systems.\cite{nakamura_exp,caspar_exp,lukens_exp,nakamura_rabi_osc,han_nature}

Effective measurement of quantum bits is a crucial step in quantum
computing. An ideal measurement of the qubit is a projective
measurement\cite{measurement_process} that correlates each state
of the quantum bit with a macroscopically resolvable state. In
practice, it is often hard to design an experiment that can both
projectively measure a solid-state qubit effectively and meanwhile
does {\bf not} couple environmental noise to the qubit. Often in
solid-state systems, the detector is fabricated onto the same chip
as the qubit and couples with qubit all the time. On the one hand,
noise should not be introduced to the qubit via the coupling with
the detector. This requires that the detector is a quantum system
well-isolated from the environment. On the other hand, to
correlate the qubit states to macroscopically resolvable states of
the detector, the detector should behave as classical system that
has strong interaction with the environment, and at the same time
interacts with the qubit strongly. These two aspects contradict
each other, hence measurements on solid-state quantum bits are
often limited by the trade-off between these two aspects.
\cite{devoret_schoelkopf_nature,schoelkopf_rfset}  In the first
experiment on the superconducting persistent-current qubit
(pc-qubit)\cite{caspar_exp},  the detector is an under-damped dc
SQUID that is well-isolated from the environment and behaves
quantum mechanically. The detected quantity of the qubit---the
self-induced flux, is small compared with the width of the
detector's wave packet.  As a result, the detector has very bad
resolution on the qubit states. This is one of the major problems
in the study of the flux-based persistent-current qubits.
\cite{pc_qubit_s,pc_qubit_l,caspar_exp}

Various attempts have been made to solve the measurement problem
\cite{averin_measurement,greenberg_measurement,milburn_measurement}
and to provide proposals for scalable quantum computer with
superconducting qubits\cite{You_scalable_charge_qubit}. In a
recent experiment on the flux qubit, the measurement efficiency
has been greatly improved by optimizing the bias current and
coherence oscillation has been observed\cite{pc_qubit_exp_2}. In
this paper, we present a new scheme that effectively measures the
pc-qubit by an on-chip detector in a ``single-shot'' measurement
and does {\bf not} induce extra noise to the qubit until the
measurement is switched on. The idea is to make a switchable
measurement (but a fixed detector) that only induces decoherence
during the measurement. During regular qubit operation, although
the qubit and the detector are coupled, the detector stays in its
ground state and only induces an overall random phase to the
qubit. The measurement process is then switched on  by resonant
microwave pulses. First we maximally entangle the qubit coherently
to a supplementary quantum system. Then we measure the
supplementary system to obtain the qubit's information. This
approach of exploiting conditional resonant transitions for signal
amplifying is different from other approaches.

In the following discussion, we illustrate this method
by applying it to the measurement of the  superconducting
persistent-current qubit (pc-qubit).  In the previous
experiment\cite{caspar_exp}, the qubit is inductively coupled to
the detector---a dc SQUID. The flux of the qubit affects the
critical current of the SQUID by an offset $\Delta
I_c=2I_c\vert\sin{(\varphi_{ext}+\delta\varphi_q)/2}-\sin{\varphi_{ext}}\vert\sim\pm
I_c\delta\varphi_q\simeq\pm 10^{-3}I_c$, where $\varphi_{ext}$ is
the flux in the SQUID loop in units of $\Phi_0/2\pi$. This offset
is recorded by measuring the switching current distribution of the
dc SQUID, the average of which is offset by $\sim\Delta I_c$ as
well. Due to quantum fluctuation and thermal activation, the
switching current distribution has a finite width that is much
larger than $\Delta I_c$. Hence, the two qubit states result in
two switching histograms whose separation is much narrower than
the width of the histogram. As a result, the histogram is not
perfectly correlated with the qubit states and the measurement has
to be repeated many times ($10^4$ times) to derive the information
of the qubit. This problem can be overcome with our method by
using an rf SQUID to be the supplementary system. Our method is
not only closely related to the ongoing experiments of the
pc-qubit, but also brings a new idea for effectively measuring
solid-state qubits with a resonant pulse technique.

One might worry that coupling the qubit to an rf SQUID brings to
the qubit a new source of noise that couples directly to the rf
SQUID. However, in our scheme, until the measurement, the coupling
between the qubit and the rf SQUID nearly commutes with the
Hamiltonian of the rf SQUID,  and the rf SQUID stays in its ground
state. The rf SQUID behaves as a poor transmitter of the noise and
can not transfer noise to the qubit as we show below.  During the
entanglement pulse, the rf SQUID induces decoherence to the qubit
in ${\rm \mu secs}$ which is much longer than the $10\,{\rm nsec}$
duration of the entanglement pulse.

\begin{figure}
\centerline{\epsfxsize=2.5in\epsfbox{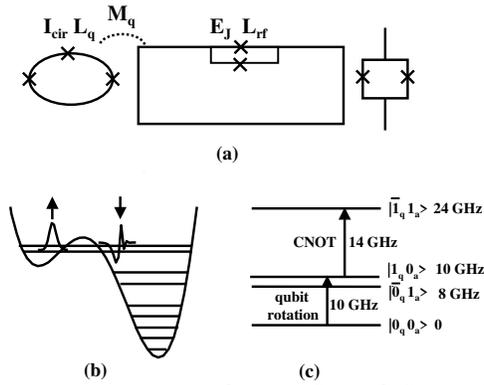}}
\caption{\narrowtext (a). Circuit of the measurement scheme,
from left to right: the qubit, the rf SQUID and the dc SQUID
magnetometer.  (b). Eigenstates and potential energy of the rf SQUID when
biased at $f_{rf}=0.4365$ flux quantum. The ETLS are labeled with arrows and
the wave functions are shown. (c). The states of the interacting
qubit and the rf SQUID.}
\label{figure_1}
\end{figure}

The superconducting persistent-current
qubit\cite{pc_qubit_s,pc_qubit_l} is a superconducting loop that
has three Josephson junctions in series, Fig.~\ref{figure_1} (a).
The qubit is controlled by the magnetic flux $f_q\Phi_0$ in the
loop, where $\Phi_0$ is the flux quantum. The qubit states of this
circuit are nearly localized flux states with opposite circulating
currents. The qubit states are analogous to the states of a $1/2$
spin and is described by the $SU(2)$ algebra of the Pauli
matrices. The qubit Hamiltonian can be written as ${\cal
H}_q=\frac{\epsilon_0}{2}\sigma_z^q+ \frac{t_0}{2}\sigma_x^q$,
where $\epsilon_0\propto (f_q-1/2)$ and $t_0$ is the coherent
tunneling between the two localized flux states over potential
energy barrier. The operator for the circulating current is
$\hat{I}_q=I_{cir}\sigma_z^q$.  Typically, the circulating current
of the qubit is $I_{cir}\approx 0.7 I_c$, where $I_c$ is the
critical current of the Josephson junctions and $I_c=200\,{\rm
nA}$. With a loop inductance of $L_q=10\,{\rm pH}$, the
self-induced flux of the qubit is $\delta\varphi_q=10^{-3}\Phi_0$.

To understand what prevents the effective measurement of the
pc-qubit in \cite{caspar_exp}, we analyzed the previous
measurement in detail in
\cite{lin_induced_noise,caspar_induced_noise}. Considering the dc
SQUID as coupled oscillators. The direct coupling between the
qubit and the dc SQUID offsets the origin of the SQUID oscillator
by $\pm\delta\varphi_0=\pi M_qI_{cir}/\Phi_0\approx 0.002$ with
typical experimental parameters, where $M_q=8\,{\rm pH}$ is the
mutual inductance between the two circuits. The overlapping
between the shifted oscillator ground states is $
\langle\psi_g^-\vert\psi_g^+\rangle=
\exp{(-\delta\varphi_0^2/2\langle \varphi_m^2\rangle)}\approx
1-0.0002$, where $\sqrt{\langle \varphi_m^2\rangle}\approx 0.1$ is
the width of the ground state wave packet of the inner oscillator
of the SQUID. The measurement of the qubit becomes the detection
and the resolution of the overlapping and highly non-orthogonal
oscillator states $\vert\psi_g^\pm\rangle$. The overlapping of the
oscillator states limits the efficiency of the previous
measurement.

Assume the qubit state is $\vert\psi_q\rangle=c_0\vert
0_q\rangle+c_1\vert 1_q\rangle$. With the inductive coupling, the
density matrix of the dc SQUID (in the previously used method of
\cite{caspar_exp}) quickly relaxes to a mixed state of
$\vert\psi_g^\pm\rangle$, $\rho_m=\vert
c_0\vert^2\vert\psi_g^+\rangle\langle\psi_g^+\vert + \vert
c_1\vert^2 \vert\psi_g^-\rangle\langle\psi_g^-\vert$. Let the
desired measurement accuracy be $A_m=x_{err}/L_x$ where $x_{err}$
is the square root error from the expected value of the measured
variable and $L_x$ is the range of the variable $x$. For a von
Neumann measurement, within $N$ measurements, the average time we
find $\vert 0\rangle$ is $\vert c_0\vert^2 N$ with the deviation
$\Delta N/N=1/(2\sqrt{N})$. $N_v=1/(2A_m)^2$ repetitions are
required to achieve the accuracy $A_m$. For a measurement with
overlapping distributions as in the previous discussions, assume
each distribution is a Gaussian for simplicity. The average of the
Gaussian functions $y_0$ and $y_1$ are slightly different, but much smaller
than $\sqrt{\sigma}$, where $\sigma$ is the deviation of the
Gaussian distributions. Given the qubit state, the average of the
measured $y$'s is $y_{ave}^{exp}=|c_0|^2y_0 + |c_1|^2y_1$, from
which we can infer $|c_0|^2$ of the qubit state. With a finite
number of measurements, $y_{ave}=\frac{1}{N}\sum_{k=1}^N y_k$ is
described by a Gaussian distribution according to the Central
Limit Theorem with an average $y_{ave}^{exp}$ and a deviation
$\sigma/N$  ($N$ measurements).  The accuracy with $N$
measurements is $\Delta
y_{ave}/|y_1-y_0|=\sqrt{\sigma/N|y_1-y_0|^2}$.
$N_p=\frac{4\sigma}{|y_1-y_0|^2}N_v$ repetitions are required to
achieve the accuracy $A_m$. In the previous experiment,
$2\sqrt{\sigma}/|y_1-y_0|=50$, so $N_p/N_v>10^3$ is required to
get satisfactory results.

This shows that the measurement in \cite{caspar_exp} is not an
efficient measurement. The quantum nature of the dc SQUID (which
is intentionally designed to reduce decoherence) prevents
efficient detection of the qubit's information. To get a more
efficient measurement, we should either encode information in the
pc-qubit in some other way or measure the qubit states with another
approach.

Given the Hamiltonian of a qubit, a projective measurement that
correlates the eigenstates to distinct macroscopic states can
always be constructed according to Neumark's
theorem\cite{measurement_process}. However, in experiments, it is
not obvious that we can build a measurement apparatus that
effectively measures the pc-qubit without introducing extra noise.
In the following, we present a new measurement scheme that
improves the previous measurement significantly and is both
effective and noiseless.

The idea is that instead of directly detecting the flux of the
qubit, we first apply a short microwave pulse to entangle the
qubit with a supplementary quantum system that behaves as an
effective two-level system (ETLS). The flux or charge of the ETLS
is designed to be much larger than that of the qubit. Then the
ETLS whose state exactly reflects the qubit state is measured.
This scheme is a highly effective ``single-shot'' measurement, and
meanwhile it avoids transferring extra noise from the detector to
the qubit.

Let the pc-qubit interact with this supplementary system via
inductive interaction $M_q\hat{I}_q\hat{I}_a$. We assume the
supplementary system is also a current loop with
$\hat{I}_a=I_{a}\sigma_z^a$ and $\sigma_z^a$ is the Pauli matrix
of the ETLS. The Hamiltonian is

\begin{equation}\label{qubit_ETLS}
{\cal H}_0=\frac{\epsilon_0}{2}\sigma_z^q +\frac{t_0}{2}\sigma_x^q
+\frac{\hbar\omega_a}{2}\sigma_z^a +\frac{t_0^a}{2}\sigma_x^a+
\frac{\hbar\omega_\Delta}{2}\sigma_z^q\sigma_z^a
\end{equation}
where $\omega_a$ is the energy splitting of the ETLS, and
$\omega_\Delta$ is the inductive interaction. We design the
tunneling $t_0^a$ to be adjustable. During qubit operation,
$t_0^a=0$; during measurement, $t_0^a$ is a resonant pulse that
flips the ETLS.   The energy levels are shown in
Fig.\ref{figure_1} (c).

During regular computation, we store the supplementary system in
its ground state $\vert 0_a\rangle$. Due to the interaction the
qubit's energy is modified as
$\hbar\omega_q=\sqrt{(\epsilon_0-\hbar\omega_\Delta)^2+t_0^2}$ and
the qubit states are $\vert
0_q\rangle=[-\sin{\frac{\theta}{2}},\cos{\frac{\theta}{2}}]^T$ and
$\vert
1_q\rangle=[\cos{\frac{\theta}{2}},\sin{\frac{\theta}{2}}]^T$ with
$\sin{\theta}=t_0/\hbar\omega_q$. By applying an external
oscillation with this frequency, single-qubit gates are achieved.
In this process, the ETLS stays in its ground state and has
trivial dynamics. When the ETLS at state $\vert 1_a\rangle$, the
qubit's energy is
$\hbar\bar{\omega}_q=\sqrt{(\epsilon_0+\hbar\omega_\Delta)^2+t_0^2}$
and the qubit states are $\vert
\bar{0}_q\rangle=[-\sin{\frac{\bar{\theta}}{2}},\cos{\frac{\bar{\theta}}{2}}]^T$
and $\vert
\bar{1}_q\rangle=[\cos{\frac{\bar{\theta}}{2}},\sin{\frac{\bar{\theta}}{2}}]^T$
with $\sin{\bar{\theta}}=t_0/\hbar\bar{\omega}_q$.

To measure the qubit's state, local operation on the supplementary
system is applied to entangle the two systems. With the presence
of the qubit, we have:
$E_{\bar{0}_q1_a}-E_{0_q0_a}=\hbar\omega_a-(\hbar\bar{\omega}_q-\hbar\omega_q)/2$
and
$E_{\bar{1}_q1_a}-E_{1_q0_a}=\hbar\omega_a+(\hbar\bar{\omega}_q-\hbar\omega_q)/2$.
By applying an external pulse of
$\frac{1}{2}\hbar\Omega_X\sigma_x^a$ in resonance with
$E_{\bar{1}_q1_a}-E_{1_q0_a}$, the ETLS is flipped to the state
$\vert 1_a\rangle$. Let $(\bar{\omega}_q-\omega_q)\gg\Omega_X$,
off-resonant transition between the states $\vert
\bar{0}_q1_a\rangle$ and $\vert 0_q0_a\rangle$ is negligible and
the dynamics only depends on the resonance properties. After a
$\pi$ pulse that operates as $\exp(i\frac{\pi}{2}\sigma_x^a)$, the
ETLS is maximally entangled with the qubit: $( c_0\vert
0_q\rangle+c_1\vert 1_q\rangle ) \vert 0_a\rangle \rightarrow
c_0\vert 0_q\, 0_a\rangle+ i c_1\vert \bar{1}_q\, 1_a\rangle$,
which gives the density matrix of the ETLS as:

\begin{equation}\begin{array}{c}
\rho_a=\vert c_0\vert^2 \vert 0_a\rangle\langle 0_a\vert +\vert
c_1\vert^2 \vert 1_a\rangle\langle 1_a\vert\\ [2mm]
+(ic_0^*c_1\langle
0_q\vert\bar{1}_q\rangle\vert 1_a\rangle\langle 0_a\vert +c.c.)
\\ [2mm] \end{array} \end{equation}
where $\langle
0_q\vert\bar{1}_q\rangle=\sin{\frac{\bar{\theta}-\theta}{2}}$. The
probabilities $\vert c_{0,1}\vert^2$ of the ETLS are then measured
by a detector. Note that the supplementary system does not have to
be a qubit and be well isolated from the environment itself. But
it is required that the noise is not transferred back to the qubit
during regular qubit operations.

In this design, we choose an rf SQUID to be the supplementary
system that inductively couples with the qubit. The circuit is
shown in Fig.~\ref{figure_1}(a). The detector is a damped dc SQUID
magnetometer. In the following, we adopt the parameters in
\cite{rf_SQUID_numbers,lukens_exp} for the rf SQUID, where
coherent manipulation of the rf SQUID has been achieved
experimentally. Typical numbers are: $L_{rf}=154\,{\rm pH}$,
$I_c=4\,{\rm \mu A}$, $C_J=40\,{\rm fF}$, and $E_J/E_C\approx
4000$. The inductance of the rf SQUID is much larger than that of
the qubit, which has two consequences: 1. the flux difference
between the states localized in the two potential wells of the rf
SQUID is of an order of half a flux quantum and can be resolved by
a dc SQUID magnetometer in a ``single-shot'' detection; 2. the
coupling between the rf SQUID and the environmental noise is
strong, hence it is harder to keep the coherence of the rf SQUID
and to use the rf SQUID as a qubit directly. At $\beta_L=2\pi
L_{rf}I_c/\Phi_0\approx 1.9$, the rf SQUID has a double-well
potential with several eigenstates localized in each well. In
practice, the junction is always made of a SQUID where
$E_J(\Phi_{ex})$ is controllable by external flux $\Phi_{ex}$ and
hence is $t_0^a$. The potential energy of the rf SQUID is drawn in
Fig.~\ref{figure_1}(b) with the energies of its eigenstates. By
adjusting the parameters, two states localized in different wells
and indicated by the up and down arrows in Fig.~\ref{figure_1}(b)
are chosen as the effective two-level system. The currents of
these two states differ by $\Delta I\approx I_c$ and results in a
flux difference of $\Delta\Phi_{rf}=\Delta IL_{rf}\approx
0.3\Phi_0$. By adjusting the flux in the qubit loop, we have
$\epsilon_0=13\,{\rm GHz}$, $t_0=1\,{\rm GHz}$, $\omega_a=11\,{\rm
GHz}$ and $t_0^a=0$. By adjusting the mutual inductance to be
$M_q/L_q=1/4$, we have $\omega_\Delta=3\,{\rm GHz}$.
The states are drawn in Fig.~\ref{figure_1} (c) with their
energies labeled beside each level.

The rf SQUID is stored in the state $\vert 0_a\rangle$ as in Fig.1
of \cite{lukens_exp} by suddenly switching the flux in the loop.
The qubit energy is $\omega_q=10\,{\rm GHz}$ and single-qubit
operation is implemented with microwave pulse at resonant
frequency. During qubit operations, the rf SQUID has trivial
dynamics. In the beginning of a measurement, a microwave pulse
with frequency $E_{\bar{1}_q1_a}-E_{1_q0_a}=14\,{\rm GHz}$ is
applied to the rf SQUID for a $\pi$ rotation. This pulse flips the
SQUID state when the qubit is in $\vert 1_q\rangle$. When the
qubit is in state $\vert 0_q\rangle$,
$E_{\bar{0}_q1_a}-E_{0_q0_a}=8\,{\rm GHz}$ and the applied pulse
is not in resonant with rf SQUID. By controlling the external flux
$\Phi_{ex}$, a $\pi$ pulse of $10\,{\rm nsec}$ ($\omega_X=50\,{\rm
MHz}$) flips the rf SQUID. The off-resonant transition has
probability $(\omega_X/\Delta\omega)^2$ which is lower than
$10^{-4}$ and is irrelevant in this process. The operation is
hence an effective controlled-not (CNOT) gate on the rf SQUID.
After the entanglement pulse, the rf SQUID is measured by a
magnetometer, such as the dc SQUID in Fig.~\ref{figure_1} (a).
Optimized designs besides the simple dc SQUID configuration can be
made for better
detection.\cite{van_duzer,TPO_book,ksegall_discussion}

In designing the rf SQUID, attention should be paid to several
issues to successfully implement this measurement scheme.  First,
the two-level system should be well separated from other states of
the rf SQUID so that no off-resonant leakage to other levels
happens during the entanglement pulse. In our design, the two
states are at least $40\,{\rm GHz}$ away from all the other states
and off-resonant transitions can be neglected. Second, a trivial
but crucial point, the parameters have to be realistic for sample
fabrication. We base our scheme on existing
experiments\cite{rf_SQUID_numbers,lukens_exp}.  Although we chose
as an example to couple the qubit inductively to an rf SQUID in
this paper, other supplementary systems can also be used with
different interaction mechanisms and different detection
technologies.

With the self-generated flux of an order of one flux quantum, the
rf SQUID is subjected to strong perturbation from the environment,
such as randomly trapped flux, impurity spins and nuclear spins.
Fortunately, however, the noise does not affect the qubit during
regular qubit operations. The flux-like noise adds to
Eq.~\ref{qubit_ETLS} a term $\sigma_z^a f(t)$ which shakes the
energy levels of the rf SQUID up and down randomly. As the ETLS stays
in its ground state $\vert 0_a\rangle$ during the qubit
operations, this term only contributes an overall phase to the
total wave function of the interacting system and does not
decohere the qubit. This is true even when the qubit Hamiltonian
has a nonzero $\sigma_x^q$ term. For environmental degrees that
assume $\sigma_x^a$ coupling with the rf SQUID, the environmental
modes with the frequency around $10\,{\rm GHz}$ can flip the rf
SQUID in principle. But at the low temperature of $20\,{\rm mK}$,
no excitations of these transversal modes exist to excite the rf
SQUID from the ground state to the excited state.

During the entanglement pulse, the rf SQUID makes a transition
from $\vert 0_a\rangle$ to $\vert 1_a\rangle$, and the flux-like
noise affects the dynamics of the qubit-ETLS system. It is
important for the gate time to be shorter than the decoherence
time of the rf SQUID to maximally entangle the qubit and the ETLS.
In the rotating frame of the Hamiltonian ${\cal H}_0$ in
Eq.~\ref{qubit_ETLS}, the Hamiltonian during the entanglement is

\begin{equation}\label{H_rot_noise}
  {\cal H}_{rot}=\frac{1+\sigma_z^q}{4}\Omega_X\sigma_x^q+f(t)\sigma_z^a
\end{equation}
the first term is the resonant pulse, and the second term is the
flux noise to the rf SQUID which dephases the rf SQUID. The noise
is treated as a classical fluctuation field $f(t)$. The
decoherence rate is determined by the spectral density $\langle
f(\Omega_X)f(-\Omega_X)\rangle$. With noise coupling as in
\cite{lin_induced_noise}, we estimate the decoherence time to be
between $0.1$ to $10\,{\rm \mu sec}$ which is significantly longer
than the resonant pulse of $10\,{\rm nsec}$.

In conclusion, we presented an effective measurement scheme that
increases the measurement efficiency without introducing extra
decoherence. We illustrated this method by applying it to the
superconducting persistent-current qubit where a supplementary
two-level system---an rf SQUID---was coupled to the
persistent-current qubit. This method improves the previous
measurement by avoiding the difficulty of measuring a flux of
$10^{-3}$ flux quantum with a quantum detector whose quantum
broadening is of $0.1$ flux quantum. Our scheme creates for the
solid-state qubits a new measurement scheme that is projective,
noiseless and switchable. More delicate designs based on this idea
can be developed to optimize the measurement.

\section*{acknowledgments}
This work  is supported in part by the AFOSR grant
F49620-01-1-0457 under the DoD University Research Initiative on
Nanotechnology  (DURINT) program and by ARDA and the ARO student
support grant DAAD19-01-1-0624.

\newpage


\begin{thebibliography}{99}
\bibitem{lin_induced_noise}
L. Tian, S. Lloyd and T.P. Orlando, Phys. Rev. B {\bf 65}, 144516 (2002);
the mass of the inner oscillator
is $m_m=2C_J(\frac{\Phi_0}{2\pi})^2$, where $C_J$ is the junction capacitance of the dc SQUID
and $\Phi_0$ is the flux quantum, the frequency is $\omega_m=\sqrt{2/L_{dc}C_J}$ where $L_{dc}$
is the inductance of the dc SQUID.
\end{thebibliography}
\end{document}